\documentclass{nature}
\usepackage{epsfig}

\title{X-ray flares following short $\gamma$-ray bursts from shock\\ 
heating of binary stellar companions}

\author{Andrew I. MacFadyen$^{1}$, Enrico Ramirez-Ruiz$^{1,3}$ \&
Weiqun Zhang$^{2,3}$}

\begin{document}

\maketitle

\begin{affiliations}
\item Institute for Advanced Study, School of Natural Sciences, Einstein Drive, Princeton,
N.J. 08540

\item Kavli Institute for Particle Astrophysics and Cosmology,
Stanford University, P.O. Box 20450, MS 29, Stanford, CA 94309

\item Chandra Fellow

\end{affiliations}

\begin{abstract}
The discovery of long-lasting ($\sim 100$ s) X-ray flares following short
gamma-ray bursts initially called into question whether they were truly
classical short-hard bursts\cite{GCN3667,GCN3570}. Opinion over the last few
years has coalesced around the view that the short-hard bursts arise from the
merger of pairs of neutron stars, or a neutron star merging with a
stellar-mass black hole\cite{eichler89,NPP92,rosswog03}. The natural
timescales associated with these processes\cite{lrrp04}, however, essentially
preclude an X-ray flare lasting $\sim 100$ s. Here we show that an
interaction between the GRB outflow and a non-compact stellar companion at a
distance of $\sim$ a light-minute provides a natural explanation for the
flares. In the model, the burst is triggered by the collapse of a neutron
star after accreting matter from the companion. This is reminiscent of type
Ia supernovae, where there is a wide distribution of delay times between
formation and explosion, leading to an association with both star-forming
galaxies and old ellipticals.
\end{abstract}

Until recently, short GRBs were known predominantly as bursts of
$\gamma$-rays, largely devoid of any observable traces at any other
wavelengths.  However, a striking development in the last several
months has been the measurement and localization of fading X-ray
signals from several short GRBs, making possible the optical and radio
detection of afterglows, which in turn enabled the identification of
host galaxies at cosmological
distances\cite{bloom05,fox05,gehrels05,xavier05}. The presence in old
stellar populations e.g., of an elliptical galaxy for GRB050724, rules
out a source uniquely associated with recent star
formation\cite{berger05}. In addition, no bright supernova is observed
to accompany short GRBs\cite{bloom05,fox05,hjorth05}, in distinction
from most nearby long-duration GRBs\cite{hjorth03}. The current view
of most researchers is that short GRBs arise when a neutron star (NS)
binary or neutron star-black hole (BH) binary, which loses orbital
angular momentum by gravitational wave emission, undergoes a
merger\cite{eichler89,NPP92,rosswog03}. Current calculations of
compact binary mergers suggest that high spatial velocities would take
these binaries, in more than half of the cases, outside of the
confines of the host galaxy before they merge and produce a burst, in
agreement with current observations\cite{bloom99,fryer99}.

Recently, long duration ($\sim 100$ s) X-ray flares have been observed
to follow several short GRBs\cite{vil05,barthelmy05}, e.g., GRB050724,
after a delay of $\sim 30$ s. There is also independent support that
X-ray emission on these timescales is detected when lightcurves of
many bursts are stacked\cite{lrg01}. One possible interpretation of
these (rapidly declining) flares is that a large fraction of energy
continues to be emitted by the GRB source for as long as
minutes\cite{barthelmy05}. This hints at the desirability for a
"central engine'' lasting much longer than a typical dynamical time
scale for a stellar mass compact object. This is in disagreement with
theoretical estimates, which suggest that NS-NS and NS-BH mergers will
lead to brief energy input episodes, typically of the order of the
duration of a short burst\cite{lrrp04}.  It is argued here that the
flare emission is, however, naturally produced by the interaction of
the extended - possibly magnetically dominated - ejecta emanating from
a short GRB with the envelope of a giant companion star. This more
isotropic GRB ejecta component need not necessarily dominate the total
burst energetics, but it can be efficiently reprocessed by the
envelope of the companion star at distances $\sim 10^{12}$ cm, into an
X-ray flare with a luminosity and timescale comparable to the observed
values. The portion of the ejecta shell not impacting the star expands
undisturbed to large radius ($10^{14}-10^{16}$ cm) where internal
shocks or magnetic dissipation convert its energy into $\gamma$-ray
photons as envisioned in standard models\cite{piranrev}. The observer
is envisioned here to be located out of the plane of the binary.  Soft
X-rays produced during the collision, although produced at smaller
radii than the $\gamma$-rays, are observed to arrive after the
GRB. Under this interpretation, the GRB engine is not required to
operate for longer than a typical burst duration of 0.1-1 s. We note
here that, in addition to the rapidly declining $\sim 100$s flare, GRB
050724 rebrightens\cite{barthelmy05} after $\sim 10^4$ s. Given its
slower decline, this feature has a satisfactory
standard\cite{piranrev} explanation: as the leading edge of fast
ejecta moves farther away from the central engine, it starts to
decelerate and is caught up by slower-moving material, creating a
'refreshed shock'.

Three snapshots from numerical simulations of a GRB ejecta shell as it
expands, makes contact with a red giant star, and ultimately engulfs
it, are shown in Figure 1.  The interaction begins with the formation
of a strong shock as the shell is rapidly decelerated from $\Gamma_i
\sim 400$ to $\Gamma_f \sim 1$ in the surface layers of the star,
dissipating a large fraction of its kinetic energy. Internal energy is
created rapidly and remains at a few $10^{49}$ erg for at least $\sim
100$ s (Figure 2).  The initial rise in internal energy takes place
immediately after the shell reaches the stellar surface, 30 s after
the GRB.  A strong shock forms, starting where the two spheres first
touch, then subsequently spreads as an increasingly larger portion of
the stellar surface is hit. The rate of energy dissipation as the
shell sweeps across the star - for a GRB located approximately one
stellar radius from the stellar surface - is roughly given by $\dot{E}
\approx E c / (4R_\ast) = 7.5 \times 10^{48} (E/10^{51}\;{\rm
erg})(R_\ast/10^{12}\;{\rm cm})^{-1}\;{\rm erg/s}$, where $E$ is the
total energy in the ejected shell, $c$ is the speed of light and
$R_\ast$ is the stellar radius. This simple estimate yields
luminosities that are consistent with those found in numerical
calculations shown in Figure 2.  A few hundred seconds after the GRB
(bottom panel of Figure 1), the shell has almost entirely wrapped
around the star and fresh material is no longer being shocked. The
internal energy declines rapidly after the GRB flare finishes shocking
the star bringing the flare emission to an abrupt end.  The decay
slope of the corresponding lightcurve may be substantially modified by
radiative effects.  However, the dissipated energy has the correct
magnitude, delay and duration to account for the observed properties
of X-ray flares in short GRBs.

The flares observed after the GRB 050724 and GRB 050709 bursts
corresponds, for their assumed distances of $z=0.257$ and $z=0.16$, to
$E_f \sim 6 \times 10^{49}$ erg and $E_f \sim 3 \times 10^{49}$ erg
respectively in the $2-25$ keV band.  How is energy dissipated during
the GRB-star interaction (Figure 2) transformed into soft X-ray
radiation? One possibility is that the GRB ejecta shell develops a
stand-off shock before encountering the stellar envelope.  The
post-shock region generates turbulent magnetic fields and accelerates
electrons which produce a synchrotron power-law radiation
spectrum. The magnetic field strength would be of order $10^4$ G at
$10^{12}$ cm, strong enough to ensure that the shock-accelerated
electrons cool promptly, yielding a power-law continuum extending into
the X-ray band. Some of these X-rays would be deflected along the
stellar surface before escaping, but about half (the exact proportion
depending on the geometry and flow pattern) would irradiate the
material in the stellar envelope. For the high radiative efficiencies
expected in relativistic shocks, radiative heating of the shocked
material would be comparable to that of bulk heating\cite{RM01}.
However, radiative heating deposits energy near the stellar surface in
layers with modest scattering optical depth, the temperature being
determined by photoionization equilibrium. This shallow radiatively
heated layer would be expected to be substantially hotter than a
deeper bulk-heated region.  For low radiative efficiencies, on the
other hand, energy deposition by bulk heating would spread over a
highly optically thick layer. In this case, the cooling rate, mainly
due to bremsstrahlung, recombination and Comptonization, would be high
enough to reduce the temperature of the bulk-heated electrons to the
equivalent black body temperatures of $T \sim$ a few tens of keV. This
suggests that, for the conditions envisaged here, most of the flare
luminosity in the X-ray band could be thermal.  This is consistent
with numerical results where a pressure of $10^{15}$ dyne cm$^{-2}$
behind the shock (see the middle panel of Figure 1), corresponds to a
black-body temperature of $\sim 2.5 \times 10^7$ K and a
characteristic photon energy of $\sim$a few keV.  We note that a
radiatively-heated layer with a density up to $n_e\sim 10^{21} \, {\rm
cm}^{-3}$, could produce strong Fe line emission provided that the
ionization parameter $\xi=\beta L/(r^2n_e)$ exceeds\cite{brr01}
$10^{3}$, where $L$ is the total luminosity of the GRB outflow and
$\beta$ is the fraction of the power that goes into X-ray continuum.
This condition is indeed satisfied unless $\beta < 10^{-2}$.  Thermal
X-ray emission could also display line features, and such signatures
should certainly be looked for.  We note that the GRB shell must be
optically thin when its radius is $\sim 10^{12}$ cm so that the
shocked star is visible through the GRB shell.  This requires that the
GRB ejecta contain $10^{-8}$ M$_\odot$ or less and suggests that a
magnetically dominated flow may be preferred\cite{usov92}.

Shock heating of a binary stellar companion has other interesting
consequences.  Before the passage of the GRB ejecta, $\sim 10^{-2}$
M$_\odot$ of the envelope of the companion star is compressed by the
shock and heated.  The deposited energy, a few $10^{48}$ erg, will
cause the outer layers of the star to expand explosively.  This will
not, however, produce a supernova, for two reasons. First, the shock
temperatures are too low for radioactive elements such as Ni$^{56}$ to
be produced. Second, the amount of ejected debris is small. Instead,
after the expanded envelope becomes optically thin, a faint
infrared/optical transient would appear a few weeks after the GRB.

Finally, we offer a few comments on the progenitor systems in which
this interaction can naturally occur. Models in which the short GRB
results from the collapse of a rapidly rotating neutron star in a
close binary system provide a natural scenario.  In the model, the
neutron star accretes matter from the stellar companion, eventually
collapsing to form a black hole. The angular momentum in the
equatorial region of the rapidly-spinning system is too large to be
swallowed immediately when the black hole forms.  The expected
outcome, after a few milliseconds, would therefore be a spinning black
hole orbited by a torus of neutron-density matter. The mass in these
disks is found to range from\cite{shapiro} $M_t\sim 10^{-3} - 10^{-2}
M_{\odot}$. If magnetic fields anchored in the disk do not thread the
black hole, then a relativistic outflow powered by torus accretion can
at most carry the gravitational binding energy of the torus. The
extractable energy in this case is several times $10^{50}\ \epsilon
(M_{\rm t}/10^{-3} M_\odot)$ ergs, where $\epsilon$ is the efficiency
in converting gravitational energy into relativistic outflow.  If
magnetic fields of comparable strength thread the black hole, its
rotational energy offers an extra (and even larger) source of energy
that can in principle be extracted via the Blandford-Znajek
mechanism\cite{BZ}.

Not surprisingly, there is more than one way to produce a rapidly
rotating, neutron star in a close binary\cite{fryer99} including, for
example, a common envelope evolution in which a neutron star is
enveloped by the expanding atmosphere of a giant
companion. Alternatively, a binary initially containing two massive
stars, could form a neutron star and helium star binary. During the
common envelope phase, the neutron star may accrete over one solar
mass and collapse\cite{bb98}. Thus, we expect these systems to be
associated with star forming regions. Reminiscent of what is observed
in type Ia supernovae, the model predicts a wide distribution of delay
times between formation and explosion, and in turn the association of
short hard GRBs with both star forming galaxies and with ellipticals
dominated by old stellar populations\cite{xavier05}.

Neutron stars inspiralling into a stellar envelope can accrete at
rates vastly exceeding the Eddington limit if the flow develops
temperatures high enough to allow neutrinos to radiate the
gravitational binding energy. The fate of the neutron star depends on
a sensitive balance between the rate at which it accretes and the rate
at which energy is deposited into the common
envelope\cite{armitage}. Its chances for survival are therefore
diminished both by increasing the accretion rate and by augmenting the
epoch of common-envelope evolution. Observationally, several binary
pulsars are known whose properties are consistent with the neutron
star having survived a phase of common-envelope evolution. Camilo et
al.\cite{camilo} identify four pulsars which have relatively large
companion masses in excess of $0.45 M_\odot$. These systems are likely
to have undergone deep common-envelope evolution with low-mass
companions\cite{heuvel} ($1-3 M_\odot$). If the neutron star is to be
able to accrete a large mass during inspiral, the initial distance
between the two stars after the first mass transfer exchange should
not be much larger\cite{bb98} than $R_*$. This ensures that a
companion star struck by the expanding GRB outflow will intercept a
large fraction of the ejected material, thus ensuring the production
of a bright flare.

Much progress has been made in understanding the nature of
cosmological gamma-ray bursts.  Still, various alternative ways of
triggering the explosions responsible for short GRBs remain: NS-NS or
NS-BH binary mergers, spun-down supramassive NS\cite{VS99} and
accretion induced collapse of a NS. The presence of X-ray flares may
help distinguish between viable progenitors. In the absence of a
supernova-like feature, the interaction of GRB ejecta with a stellar
binary companion may be the only observable signature in the
foreseeable future shedding light on the identity of the progenitors
of cosmological short bursts.

\bibliography{mnemonic,shb}

\begin{addendum}
 \item Conversations with J. Bloom, A. Gruzinov, P. Hut, W. Lee, and
 D. Uzdensky are gratefully acknowledged. We thank N. Gehrels and
 D. Pooley for sharing unpublished {\it Swift} data on GRB 050724. AIM
 acknowledges support from the Keck Fellowship at the Institute for
 Advanced Study.  ERR and WZ acknowledge support from NASA through the
 Chandra fellowship program.  Part of this work was supported by NASA.
 Computations were performed with the Scheides Beowulf cluster at the
 Institute for Advanced Study.  The software used in this work was in
 part developed by the DOE-supported ASC / Alliance Center for
 Astrophysical Thermonuclear Flashes at the University of Chicago.
 Specifically, the RAM code\cite{ZM05} utilizes the PARAMESH AMR and
 the IO tools from FLASH2.3.

 \item[Competing Interests] The authors declare that they have no
competing financial interests.
 \item[Correspondence] Correspondence and requests for materials
should be addressed to A.~I.~M.~(aim@ias.edu), E.~R.~(enrico@ias.edu)
or W.~Z.~(wqzhang@slac.stanford.edu).
\end{addendum}


\clearpage

\begin{figure}
\centerline{ \epsfig{file=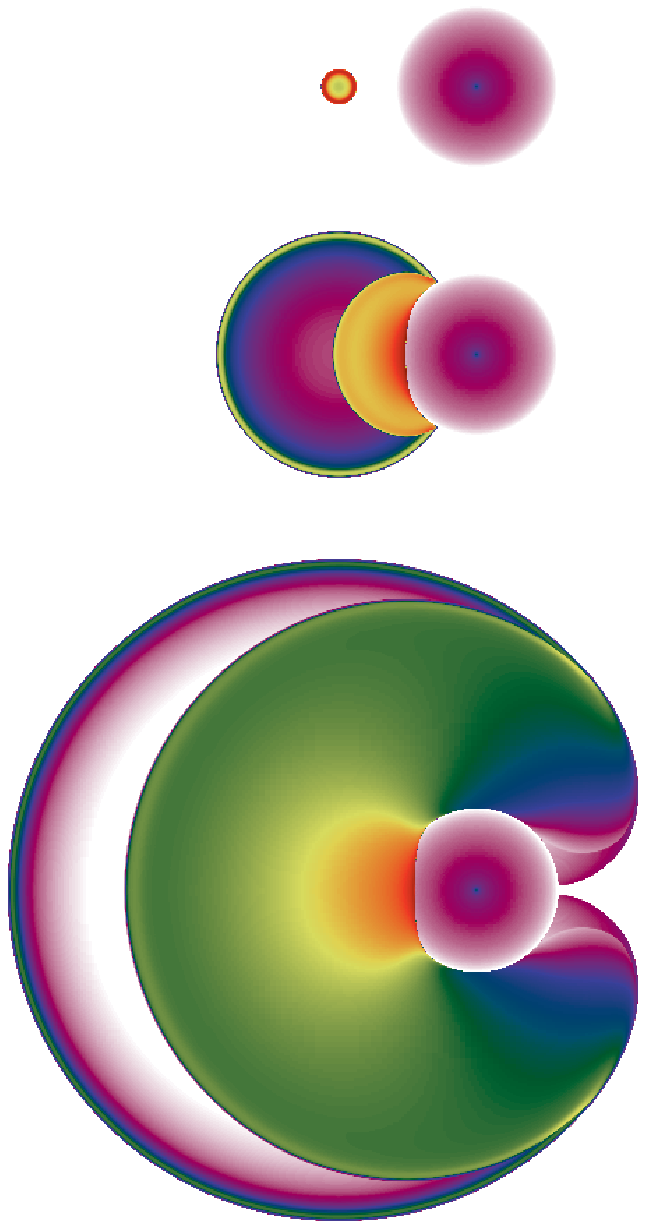,width=0.8\textwidth}}
\caption{

  \small{Collision of GRB ejecta with a companion red giant star
    showing the logarithm of pressure at $t=10 s$ (top), $t=42 s$
    (middle) and $t=184 s$ after the GRB .  The observer is located
    envisionedo be out of the binary plane toward the top of the
    figure.  In the top panel, the small orange circle is the GRB
    ejecta and the larger purple sphere is a red giant star with
    radius $1.5 \times 10^{12}$ cm.  The ejecta form a spherical shell
    containing $2 \times 10^{25}$ g ($10^{-8} M_{\odot}$) expanding at
    nearly the speed of light.  30 seconds after the GRB, the ejecta
    collide with the surface of the red giant and are rapidly
    decelerated in a strong shock. The internal energy produced during
    the collision is capable of explaining the energetics of the
    observed X-ray flares. The middle panel shows the GRB blastwave
    and shocked star 12 seconds after impact.  The dark red region is
    the X-ray emitting region.  In the bottom panel, the blastwave has
    wrapped around the star.  At this time, the energy of the
    blastwave is mainly kinetic.  The orange region in the middle
    panel and the green region in the bottom panel show the reflected
    ejecta shell traveling leftward toward to GRB source. The
    undisturbed shell expands close to the speed of light for at least
    two orders of magnitude in radius before producing the prompt
    GRB. We used the RAM code\cite{ZM05} to solve the conservative
    equations of special relativistic hydrodynamics. The computational
    domain for the simulations was $0<r<1.2\times10^{12}$ cm with a
    maximal resolution of $\Delta r = 4.7\times10^7$ cm for 1D and
    $-1.2\times10^{13}$ cm$<z< 1.2\times10^{13}$ cm, $0<r<
    1.2\times10^{13}$ cm, with $\Delta z = \Delta r = 7.32\times10^8$
    cm for 2D.  A gamma law equation of state with $\Gamma = 4/3$ was
    used and the ambient density was $100$ baryons cm$^{-3}$.}  }
\label{fig:1}
\end{figure}

\clearpage

\begin{figure}
\epsfig{file=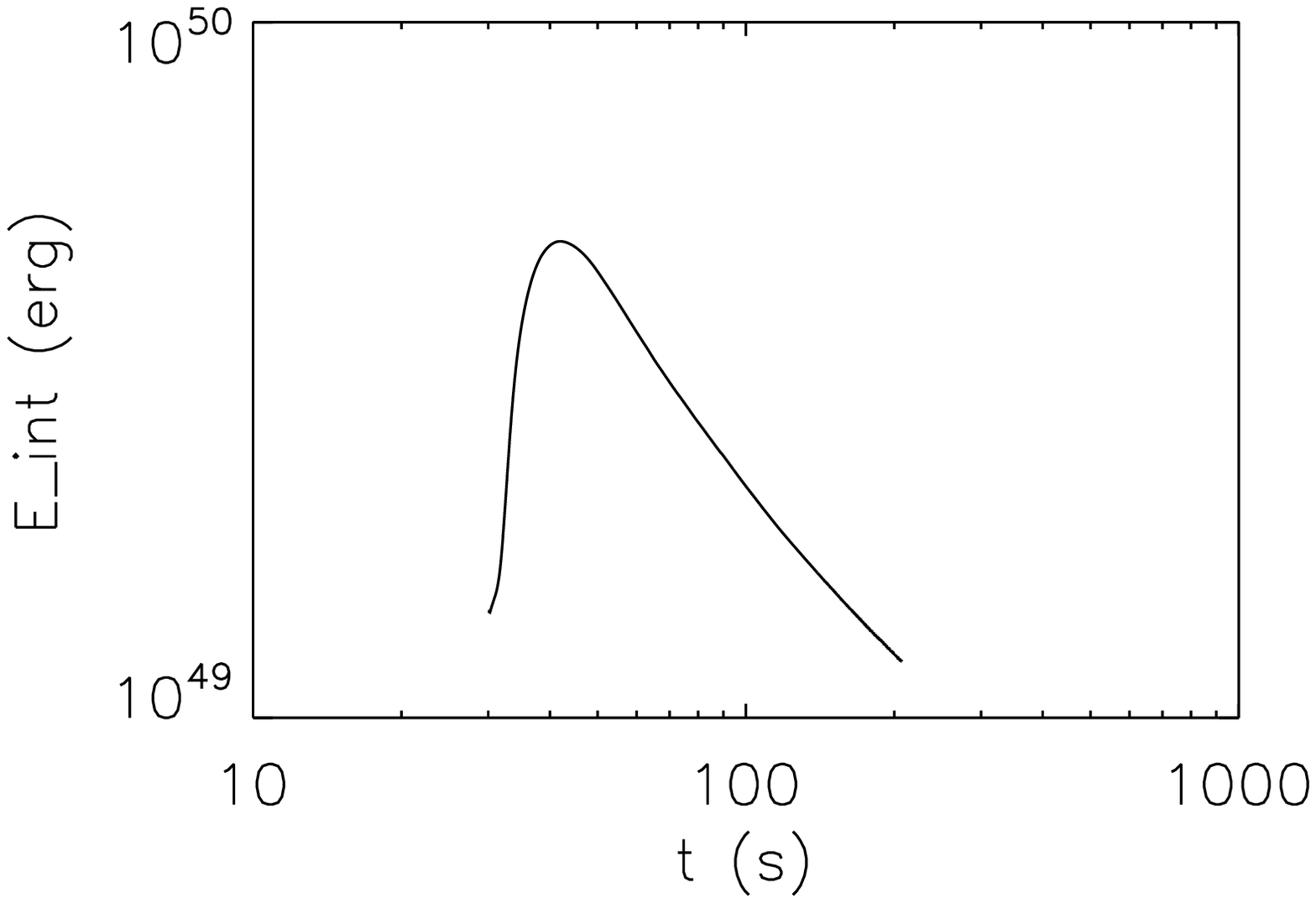,width=6 in}
\caption{\small{Total internal energy.  As the cold GRB shell collides
  with the stellar surface a large fraction of the energy in the blast
  wave is converted to internal energy in a shock.  Internal energy is
  produced with magnitude, duration and delay appropriate for being
  the source of X-ray flares observed to follow short GRBs. The sharp
  decline in the flare is easily explained by the finite time the
  companion's surface is actively shocked. In this figure, the
  remaining internal energy is being reconverted to kinetic energy and
  is not expected to contribute to the flare lightcurve.} }
\label{fig:2}
\end{figure}

\end{document}